# Fast response of pulsed laser deposited Zinc ferrite thin film as a chemo-resistive gas sensor

Saptarshi De

Department of Metallurgical Engineering and Materials Science
Indian Institute of Technology Bombay, Mumbai, India.
Email: sapjaki@gmail.com

*Abstract—* **Thin films of ZnFe₂O₄ deposited by pulsed laser technique are here demonstrated as one of the interesting materials for sensing of ethanol. The response transients were fitted well to one-site Langmuir adsorption model. Activation energies for (I) adsorption and reaction of ethanol and (II) desorption (i.e. recovery process) of ethanol from zinc ferrite thin film surface were obtained on the basis of this model. In this paper, we showed the effect of operating temperature and gas-concentration on the response time of thin film sensor materials. At the operating temperature 340ºC, the ZnFe₂O₄ thin film showed high (84%) as well as immediate response to 500 ppm of ethanol, with its resistance being saturated within ~12 seconds, which stands far superior to the response time of nano crystalline powders. Those films were also observed to have a good repeatability of their sensor response, thus representing a major step towards low-cost large-scale production of this class of devices.**

## I. Introduction

Earlier, gas sensors were made of simple oxides including $ZnO$, $SnO_2$ and $TiO_2$ [1,2,3]. They were used as sensing materials in pellets, thick films, thin film and nanowires forms [4]. But the problems with these simple oxide gas sensors are their poor selectivity and high operating temperature (400 – 550 °C). Use of mixed valence oxides and composite oxides may improve the selectivity of gas sensors in comparison to simple oxides, as the activation energies for adsorption – desorption of gas species on metal oxide surface may differ for cations. An alternative for these metal oxides are complex oxides e.g. ferrites, as complex oxides have periodic electronic structure and there has been increased interest on the use of ferrites as gas sensors. Ferrites have high resistivity which changes when the surrounding gaseous atmosphere changes [5].

Almost all ferrites, $MFe_2O_4$ (with M = Zn, Mg, Ni, Cd, Bi, Co and Cu) has been observed to have interesting gas sensing properties [6,7,8,9,10,11]. Spinel ferrite in different forms e.g. nanoparticles, thick films and thin films have been utilized for gas sensing. There are reports on gas sensors made of nanocrystalline particles, which have been synthesized by various techniques including sol gel, combustion synthesis, chemical co-precipitation method and hydrothermal method [6-8]. These powders were either directly made pellets or made as thick film by using a suitable binder and casting in substrates like glass and used for gas sensing [6-10]. Among all ferrites, only zinc ferrite was used in "resistive thin film gas sensing device" and the sensing layers were fabricated using spray pyrolysis [12,13] or spin coat technique [14] due to its low resistivity comparing to the other ferrites. Gas sensing properties should be studied with dense ferrite thin films deposited by any physical vapour deposition system (e.g. radio freequency sputtering or pulsed laser ablation), where only top surface is available for sensing. Moreover, due to its high intrinsic surface-to-volume ratio, thin film technology exhibits high capabilities for miniaturization and very short response. Here, in this report the gas sensing properties of Zinc ferrite thin films deposited by pulsed laser deposition (PLD) technique are studied.

It is seen that like other semiconducting metal oxide gas sensors, maximum effort was given to achieve promising gas sensing characteristics by the synthesis of pure phase, porous, nano-crystalline ferrite particles. The sensing mechanism was explained on the basis of reactivity with adsorbed oxygen, but the quantification of activation energy for this reaction was not done for ethanol vapor mixture on zinc ferrite surface. In this report an approach (using one-site Langmuir adsorption model) was made to find out the values of those activation energies.

## II. Experimental

Zinc ferrite (ZFO) thin films were deposited on amorphous fused silica substrate by pulsed laser deposition apparatus using a home-made sintered ceramic target of pure $ZnFe_2O_4$ with a relative density of 92%. Eximer laser (KrF) of 242 nm with 10 ns pulsed width was used to ablate the ZFO target. Laser fluence and repetition rate of the laser shots were fixed at 4 J/cm² and 10 Hz respectively. The pressure inside the deposition chamber was $5 \times 10^{-6}$ mbar or lower before deposition was commenced. During the deposition of films, the oxygen pressure inside the chamber and the target-to-substrate distance were kept at $1.3 \times 10^{-1}$ mbar and 4.5 cm respectively. At room temperature (RT) (substrate temperature $T_s= 20ºC$), $4 \times 10^4$ laser shots were used to get approximately 500 nm thick ZFO films. Thickness calibrations were performed with DEKTAK profilometer and cross section view through FEI Quanta 200 FEG-SEM. The phase and structural properties were determined by X-ray Diffraction using Pananlytical diffractometer with the $CuK^{\alpha}$ radiation ($\lambda$=1.5418 Å). Microscopic studies were carried out also with Quanta 200 SEM. Our simplified chemo-resistive gas sensor consists of a electrically insulated silica substrate, the sensitive ZFO layer and two interdigitated silver electrodes on top of the ZFO layer. After thermal annealing of ZFO thin films, 1 micron thick silver electrodes were deposited on the surface of the ZFO layer by DC sputtering technique. The electrical response of the sensing layer was investigated by Keithley 2635 source meter as a function of temperature, time, and test gas concentration. The test devices were placed on a PID controlled heater operating inside a chamber with controlled atmosphere. Resistance measurement of the sensor films was carried out with two types of gas atmospheres: dry air and a mixture of "air+test gas". Calibrated ethanol of 500 ppm (balanced with dry air) was used as test gases. Total flow of gases inside the chamber was fixed at 100 cm³/min. Further, test gas concentration was varied by mixing of dry air and the calibrated gases from two separate cylinders with the help of mass flow controller (MKS 647B). The concentration of mixed gas in the reaction chamber could be calculated using following relation

$$C_{mixed\ test\ gas} = [C_{test\ gas}(dV_{test\ gas}/dt)]/[dV_{test\ gas}/dt + dV_{dry\ air}/dt] \quad (1)$$

## III. Result and discussion

Fig. 1 shows XRD patterns of the sintered ZFO target used for thin film deposition, as-deposited ZFO thin film and the films annealed in air for 2 hours. XRD pattern of the sintered pellet matched well with single phase cubic zinc ferrite (JCPDS no. 01-089-4926). Crystalline growth was observed for the films annealed above 250°C. Average crystallite size was calculated using Scherrer formula and tabulated in table 1.

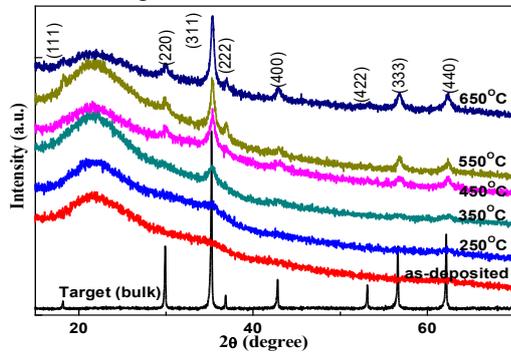

Fig. 1 XRD pattern of ZFO thin films deposited at room temperature, annealed at different temperature and target material

Table I: Crystallite size of ZFO thin films calculated using Scherrer formula

| Annealing temperature | Crystallite size |
|---|---|
| RT (as-deposited) | - |
| 250°C | - |
| 350°C | ~19nm |
| 450°C | ~25nm |
| 550°C | ~31nm |
| 650°C | ~36nm |

It is known that the gas sensing performance is better in the semiconducting ceramic oxides with finer crystallite size; therefore, in the present gas sensing study we choose the sample which was annealed at 350°C. Fig. 2 shows the surface and cross sectional view of ZFO thin film annealed at 350°C. It shows dense columnar grain growth of zinc ferrite upon thermal treatment of as deposited samples, whereas significant change in particle size along surface can't be observed from top view SEM.

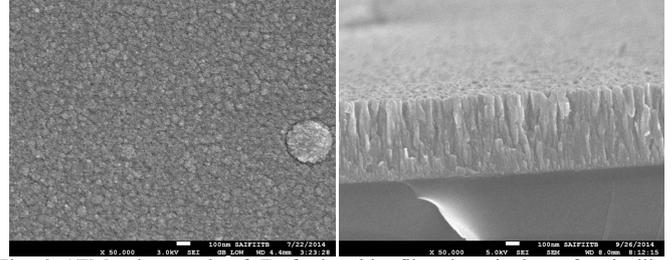

Fig. 2 SEM micrograph of Zn-ferrite thin film deposited on fused silica substrate at room temperature and annealed at 350°C, top view (left) and cross sectional view (right)

Fig. 3 shows response and recovery transient of the ZFO thin film sensor towards 500 ppm of ethanol at operating temperature of 340°C. Base line (resistance of the sensor in air) drift (~9%) was observed over long time (more than 3hr) exposure of sensing element to air and also when the ambient atmosphere of the sensor was switched back and forth between air and test gas. Whereas sensitivity (S=$\Delta R/R_{air}$ x 100%) of this sensor at 340°C varies in between 84-86% towards 500 ppm of ethanol.

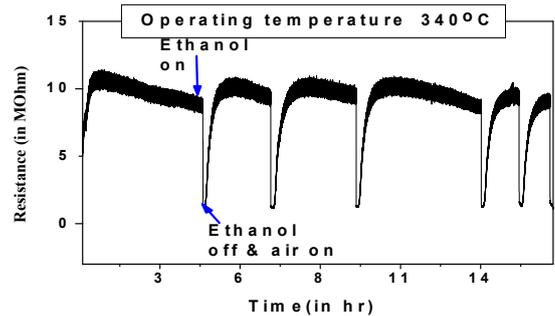

Fig. 3 Repeatability of the sensor: resistance transient of Zn-ferrite thin film gas sensor to 500 ppm of ethanol at 340°C operating temperature

As reported in several research journals, the reaction mechanism for the sensing of reducing gases by an n type Zinc ferrite semiconductor could be summarized as follows [1-14]. In a first step, at the operating temperature, oxygen is physiadsorbed in the sensor surface followed by the electron transfer from the semiconducting ferrite to adsorbed oxygen to form chemical bond between the adsorbed oxygen and the semiconducting oxides. These reactions are described in Eqs. 2 and 3 respectively

$$O_2 + sensor_{surface} \leftrightarrow O_{ad-surface} \quad (physiadsorption) \quad (2)$$
$$O_{ad-surface} + e \leftrightarrow O_{ad}^- \quad (chemiadsorption) \quad (3)$$

The exact nature of chemiadsorbed oxygen is subject to debate and depending on the temperature of adsorption oxygen may be atomic or molecular origin. When the sensor is exposed to reducing gas (R) atmosphere, the reducing gas is physiadsorbed on the sensor surface and reacts with the adsorbed oxygen according to the reaction 4 and 5 and the by-products go out (Eq.6).

R + [sensor] ↔ R$_{ad}$ (physiadsorption)       (4)
R$_{ad}$+ O$_{ad}$ ↔ R-O$_{ad}$ + e       (5)
R-O$_{ad}$ ↔ By-products$_{gas}$↑+ sensor$_{surface}$       (6)

Following the above reaction sequence, the ethanol sensing behavior of n type zinc ferrite can be envisaged as follows: as the sensing material was sintered at lower temperature (350°C), grain to grain contact has been established. Due to the chemiadsorption of oxygen (molecular or atomic anionic species) an electron depleted layer is formed at the grain surface which eventually leads to the formation of potential barrier for grain to grain electron-percolation. At temperature T, the conductance (G) of the sensor is determined by the barrier height through the well known Schottky relation [15,16]

G = G$_o$ exp(-eV$_s$/ k$_B$T)       (7)

Where eV$_s$ is the Schottky barrier and k$_B$ is the Boltzmann constant. When the sensor is exposed to ethanol, it reacts with chemiadsorbed oxygen species (see Eq.5), resulting the lowering of potential barrier which leads to the increase in conductance (or decrease in resistance). At constant T, assuming Langmuir isotherm adsorption kinetics for a single adsorption site, the conductance transient for response [G(t$_{response}$)] was given by the following equation 8.

G(t$_{response}$)= G$_o$ + G$_1$[(1 − exp(− t/ t$_{response}$)]       (8)

Where, G$_o$ is base conductance (saturated conductance with test gas) of the sensing material and τ$_{response}$ is response time. Similarly, for the recovery process the transient conductance could be expressed as

G(t$_{recovery}$)= G$_o$' + G$_1$' exp(− t/ τ$_{recovery}$).       (9)

Where, G$_o$' is saturated conductance in air and τ$_{recovery}$ is the recovery time.

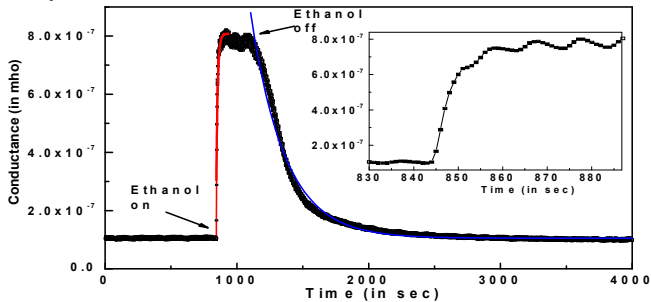

Fig. 4 Fitting of conduction transient for response and recovery of ZFO thin film sensor for 500 ppm ethanol measured at 340 °C using Langmuir adsorption model. Inset figure shows fast response time 10 sec, towards 500 ppm of ethanol measured at 340 °C.

Figure 4 shows conduction transient for response and recovery of ZFO film sensor, fitted with one site Langmuir adsorption model. The response and recovery curves were fitted well with R$^2$ value ~0.98. Inset figure (enlarged response curve) shows fast saturation of conductance within ~12 seconds for 500 ppm ethanol measured at 340°C. Small oscillation in conduction transient was due to periodic heating by PID controller. Equation 8 & 9 used to find out response and recovery time of ZFO thin film sensor operated at various temperature range (260-340°C) and ethanol concentration (5-500ppm) for further calculations.

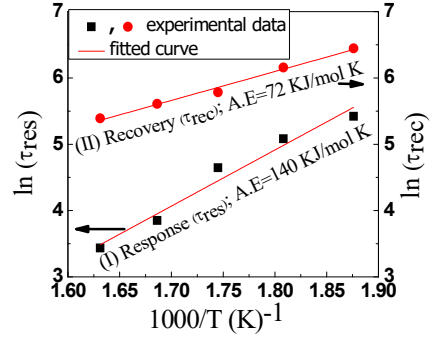

Fig. 5 Variation of response or recovery time (τ) with the operating temperature at fixed ethanol concentration (500 ppm) of Zn-ferrite thin film gas sensor.

The response and recovery time obtained from response transient using eq. 8 and 9 measured at various temperature (at fixed concentration, 500 ppm ethanol) plotted in figure 5. It is observed that response and recovery time was inversely proportional to the operating temperature. When the kinetics of sensor response is controlled by adsorption/desorption process the response time (τ) constant usually follow the following temperature dependence equations [16].

τ = τ$_o$ exp(E$_A$/2kT)   or   τ = τ$_o$ exp(E$_D$/2kT)       (10)

Where E$_A$ is the activation energy (A.E) for the chemi-adsorption followed by surface reaction (with oxygen) of ethanol and E$_D$ is the activation energy during the recovery process. As predicted by the above equation, a linear fit was obtained when ln τ was plotted with inverse of temperature. From the slope of the linear fit, the activation energy for adsorption of ethanol was found to be 1.46 eV (140 kJ/mol K), and for recovery it was 0.75 eV (72 kJ/mol K). This type of fittings to obtain the values of activation energies towards ethanol is also not available for metal oxide systems. Whereas the reported values of E$_A$ and E$_D$ are 0.27 eV and 0.42 eV respectively for hydrogen, from the combination of two site model used in nano-crystalline magnesium-zinc ferrite powders [17]; and for the same system these (E$_A$ and E$_D$) values are 1.03 and 0.19 eV for CO gas. Similarly for nano-crystalline zinc ferrite powder, E$_A$ and E$_D$ are reported as 0.56 and 0.75 eV respectively for H$_2$ [16]. For pure magnesium ferrite nano powders; E$_A$ and E$_D$ are reported as 1.45 and 0.51 eV respectively for H$_2$, and 0.61 and 0.47 eV respectively for CO [18]. From the figure 4.20, it is clear that the 'response time' or 'recovery time' will be shorter means the adsorption or desorption process will be faster with rising operating temperature in the case of high A.E.

'Response time' (τ$_{res}$) obtained by the fitting of response curve (eq. 8) which was recorded at different ethanol and hydrogen (in addition) gas concentration is plotted in figure 6. For higher than 50 ppm of gas concentration, the 'response time' seems to be less dependent on the gas-concentration. An explanation could be given as, saturation of response time (τ$_{res}$) at higher test-gas concentration may be due to reduction of available reactive sites (pre adsorbed oxygen at active sites on this sensor-surface). So, below 50 ppm of gas concentration, the variation in $\tau_{res}$ is fitted by power law of time constant [$\tau = \tau_o C_{gas}^{-\beta'}$] [not shown in the figure], and $\beta'$ was found 0.74 towards ethanol for zinc ferrite thin film sensor.

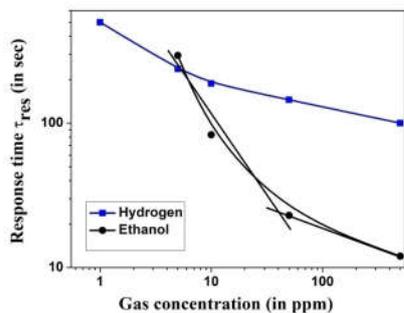

Fig. 6 Variation of response time ($\tau_{res}$) of ZFO thin film gas sensor with various ethanol and $H_2$ concentration, measured at 340 °C.

Similarly for the $H_2$ gas, value of $\beta'$ was obtained as 0.32. This type of fittings (assuming single site) with ethanol and hydrogen is not available for ferrite systems. Mukherjee et al. used two sites model for nano-crystalline zinc ferrite powder towards $H_2$ gas and the value of $\beta'$ for each site was ~0.43 [1615]. They also observed a tending to saturation in 'site 2' at higher gas concentration (> 500 ppm). Similarly, they have reported the value of $\beta'$ for magnesium ferrite nano powder as 0.48 and 0.60 for $H_2$ and CO respectively [18]. The derivation of this power law ["response time vs gas-concentration"] is not available in the literature. Mukherjee et al. [1615,18] used this empirical equation to characterize the sensor material.

In conclusion, a single-site gas adsorption model is enough to fit the response transients of the zinc ferrite thin films whereas double-site model is required for the 'sensor-pellet' made from nano crystalline zinc ferrite powder [16]. The thin film deposited by PLD technique has the shortest response time (~12 s) towards reducing gas (ethanol) in comparison to the zinc ferrite thin film deposited by other techniques i.e. spray pyrolysis and spin coat [13-14]. An indirect method was used through the kinetic analyses of the conductance transients during response and recovery, we have estimated the activation energies for relevant gas adsorption ($E_A$) and desorption ($E_D$) during ethanol sensing.

## Acknowledgment

The above research work was partially supported by the research grant from DST-ANR research project (Project Code.14IFCPAR001). We would like to thank IRCC, IIT Bombay for BDS Facility and SAIF, IIT Bombay for SEM of the samples.